\documentclass[a4paper,10pt]{iopart}
\usepackage{graphicx}

\usepackage{amssymb}
\newcommand{\trr}{\triangle}
\newcommand{\sqq}{\square}
\newcommand{\bk}{\mathcal{B}}
\newcommand{\rh}{\mathcal{R}}
\newcommand{\ptt}{\mathcal{P}_{\triangle}}
\newcommand{\ps}{\mathcal{P}_{\square}}

\begin{document}

\letter{Phase behavior of hard spheres confined between parallel hard plates: Manipulation of colloidal crystal structures by confinement}

\author{Andrea Fortini and Marjolein Dijkstra}
\address{Soft Condensed Matter, Utrecht University, Princetonplein 5, 3584 CC Utrecht, The Netherlands.}
\ead{a.fortini@phys.uu.nl}

\begin{abstract}
We study the phase behavior of hard spheres confined between two
parallel hard plates using extensive computer simulations. We
determine the full equilibrium phase diagram for arbitrary
densities and plate separations from one to five hard-sphere
diameters using free energy calculations. We find a first-order
fluid-solid transition, which corresponds to either capillary
freezing or melting depending on the plate separation. The
 coexisting solid phase consists of crystalline layers with
either triangular ($\trr$) or square ($\sqq$) symmetry. Increasing
the plate separation, we find a sequence of crystal structures
from  $\cdots n \trr \rightarrow (n+1) \sqq \rightarrow (n+1) \trr
\cdots$, where $n$ is the number of crystal layers, in agreement
with experiments on colloids. At high densities, the transition
between square to triangular phases are intervened by intermediate
structures, e.g., prism, buckled, and rhombic phases.
\end{abstract}

\maketitle

\section{Introduction}
The physics of confined systems is important in different fields
of modern technology, like lubrication, adhesion and
nanotechnology. The study of simple models is instrumental in
understanding the behavior of complex systems. As such the
hard-sphere system plays an important role in statistical physics;
it serves as a reference system for determining the structure and
phase behavior of complex fluids, both in theory and simulations.
The bulk phase behavior of hard spheres is now well understood.
At sufficiently high densities, the spheres can maximize their
entropy by forming an ordered crystal phase
\cite{Hoover1968,Pusey1986}.
The insertion of a hard wall in such a fluid decreases the number
of hard-sphere configurations. The system can increase its entropy
by the spontaneous formation of crystalline layers with triangular
symmetry, the (111) plane, at the wall, while the bulk is still a
fluid \cite{Dijkstra2004}.
This effect is known as prefreezing, and is
analogous to complete wetting by fluids at solid substrates. It is
induced by the presence of a single wall and should not be
confused with capillary freezing. Capillary freezing denotes the
phenomenon of confinement induced freezing of the whole fluid in
the pore  at thermodynamic state points where the bulk is still a
fluid. This transition  depends strongly on the plate separation.
The opposite phenomenon, called capillary melting, can also occur.
The capillary induces melting for thermodynamic state points that
correspond to a crystal in the bulk.
Confinement can also change dramatically the equilibrium crystal
structure. In 1983 Pieranski \cite{Pieranski1983} reported a
sequence of layered solid structures with triangular and square
symmetry for colloidal hard spheres confined in a wedge. The
sequence of high density structures is determined more accurately
in recent experiments \cite{Neser1997,Fontecha2005}, reporting the
observation of prism phases with both square and triangular
symmetry. Recently Cohen \cite{Cohen2004} studied configurations
of confined hard spheres under shear, demonstrating the importance
of the equilibrium configurations in the rheological properties.
Despite the great number of theoretical and simulation studies on
confined hard spheres \cite{Schmidt1996,Zangi2000,Messina2003},
the full {\em equilibrium} phase behavior  is yet unknown. In
fact, many of the previous studies were based on an order
parameter analysis, which fails dramatically in discriminating the
different structures at high densities and large plate
separations. More importantly, free energy calculations of
confined hard spheres are prohibited so far due to the lack of  an
efficient thermodynamic integration path which relates the  free
energy of interest to that of a reference system, while a further
complication arises from the enormous number of possible solid
phases that has to be considered. Hence, it is unresolved whether
the experimentally observed phases are stabilized kinetically or
are thermodynamically stable.
\begin{figure}[htb]
  \begin{center}
  \begin{indented} 
\item[] \includegraphics[width=10cm]{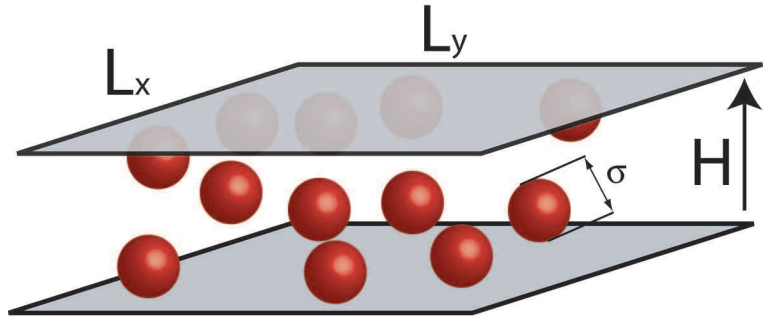}
\end{indented}
    \caption{(Color online) Illustration of hard spheres with diameter $\sigma$, confined
between parallel hard plates of area $A=L_x L_y$ and a separation
distance $H$. }
    \label{fig:model}
  \end{center}
\end{figure}

In this letter, we present a novel efficient thermodynamic
integration path that enable us  to calculate the free energy of
{\em densely packed and confined} hard spheres, with high accuracy
close to the fluid-solid transition. This method allow us  to
determine for the first time the stability of the structures found
in experiments. To this end we perform explicit free energy
calculations to map out the full phase diagram for
plate separations from 1 to 5 hard-sphere diameters. We report a
dazzling number of {\em thermodynamically stable} crystal
structures (26!) including triangular, square, buckling, rhombic,
and prism phases, and a cascade of corresponding solid-solid
transformations. In addition, the free energy calculations allow
us to determine the chemical potential at coexistence, that was
unaccessible in previous simulations. From the analysis of the
chemical potential, we find an intriguing sequence of capillary
freezing and melting transitions coupled to a structural phase
transition of the confined crystal.  We note that our new method
and  results are also relevant for confined simple
fluids~\cite{Klein1995,Gao1997,Ghatak2001} and self-assembled
biological systems~\cite{Ciach2004}. In addition, the structure of
dense packings of spheres explains the shape of for instance
snowflakes, bee honeycombs, and foams, and it is of great
importance for fundamental research, e.g., solid state physics and
crystallography, and  for applications like communication science
or powder technology \cite{book}.

\section{Model and Method}
Our model system consists of $N$ hard spheres with  diameter
$\sigma$, confined between two parallel hard plates of area $A=L_x
L_y$ (Fig. \ref{fig:model}).  In each layer we used approximately 200 particles. We use the packing fraction
$\eta=\pi\sigma^3 N/(6 AH)$ as a dimensionless density, where $H$
is the distance between  two plates.
We determine the equilibrium phase diagram by performing Monte
Carlo (MC) simulations in a box, which is allowed to change its
shape to accommodate different types of crystals; the ratio
$L_x/L_y$ may vary while $H$ and $A$ are fixed. Trial solid
structures are obtained from crystals with triangular or square
symmetry relaxed with MC moves while slowly increasing the density
by expanding the spheres. The free energy $F$ for the resulting
equilibrated structures is calculated as a function of $\eta$ and
$H$. We use the standard thermodynamic integration technique
\cite{Frenkel2002,Fortini2005a}, but with a new and efficient path
based  on penetrable potentials, that enable us to change
gradually from a non-interacting system to the confined hard-core
system of interest. The sphere-sphere potential reads
\begin{equation}
v_{ij}(R_{ij})= \left \{ \begin{array}{ll}
 \epsilon \exp(-{\rm A} R_{ij} ) & \textrm{ if $R_{ij} < \sigma_c$ } \\
0 & \textrm{ otherwise}
\end{array} \right. \ ,
\label{eq:hs}
\end{equation}
and the wall-fluid potential
\begin{equation}
v_{wi}(z)= \left \{ \begin{array}{ll}
 \epsilon \exp(-{\rm B} z_i ) & \textrm{ if $z_i < \sigma_c/2$ } \\
0 & \textrm{ otherwise}
\end{array} \right . \ ,
\label{eq:hs}
\end{equation}
where $R_{ij}$ is the distance between spheres $i$ and $j$, $z_i$
is the distance of sphere $i$ to the nearest wall, A and B are
adjustable parameters that are kept fixed during the simulations,
and $\epsilon$ is the integration parameter. The limit $\epsilon
\rightarrow \infty$ yields the hard-core interaction, but
convergence of the thermodynamic integration is already obtained
for $\epsilon \sim 70 k_B T$. The reference states ( $\epsilon = 0
k_B T$) are the ideal gas and the Einstein crystal for the fluid
and solid phase, respectively. We use a 21-point Gaussian
quadrature for the numerical integrations and the ensemble
averages are calculated from runs with 40000 MC cycles (attempts
to displace each particle once), after first equilibrating the
system during 20000 MC cycles. We  determine phase coexistence by
equating the grand potentials $\Omega=F-\mu N$ \cite{evans}.
\begin{figure}
  \begin{center}
     \begin{indented} 
\item[]\includegraphics[width=10cm]{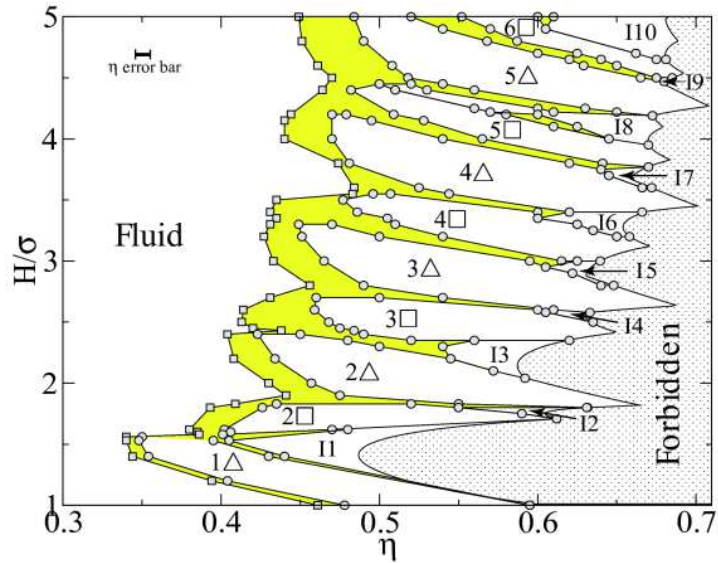}
\end{indented} 
  \caption{(Color online) The equilibrium phase diagram of hard spheres with diameter $\sigma$ confined between two parallel hard
    walls in the plate separation $H$ - packing fraction $\eta$ representation.
    The white, yellow and dotted regions indicate the stable one-phase region, the two-phase coexistence region, and the forbidden region, respectively. }
        \end{center}
\label{F:phd1}
\end{figure}

\begin{table}
\caption{List of intermediate structures $In$ as found in our
simulations and in the experiments of Fontecha
\cite{Fontecha2005}.}
\begin{indented} 
\item[]\begin{tabular}{llll}
\br
Phase & Transition & Simulation\footnote{In parenthesis metastable phases} & Experiment \\
\mr
I1 &$1\trr \rightarrow 2\sqq$& $2\bk$&$2\bk$ \\
I2 & $2\sqq \rightarrow 2\trr$&$2\rh$& $2\rh$\\
I3 &$2\trr \rightarrow 3\sqq$ &$2\ptt + 2\ps$&$2\ptt + 2\ps$ \\
I4 & $3\sqq \rightarrow 3\trr$&$3\rh +3\ps$ +(3$\bk$)&$3\rh +3\ps+ 3\ptt$\\
I5 & $3\trr \rightarrow 4\sqq$&$3\ptt +4\bk$&$3\ps$\\
I6 &$4\sqq \rightarrow 4\trr$& $4\ps +4\rh+4\ptt$ &$4\ps +4\ptt$\\
I7& $4\trr \rightarrow 5\sqq$&$4\ptt$&$4\ptt$, $4\ps$, $H$\footnote{Not fully characterized structure with hexagonal symmetry}\\
I8&$5\sqq \rightarrow 5\trr$& $5\ps$+$4\ptt$+($5\ptt$)+$5\rh$&$5\mathcal{P}$\footnote{Not fully characterized prism structure.}\\
I9&$5\trr \rightarrow 6\sqq$ & $5\ptt$ & no data  \\
I10&$6\sqq \rightarrow 6\trr$ &$5\ps+5\ptt$& no data \\
\mr
 \end{tabular}
 \end{indented}
\label{tab:list}
 \end{table}

\section{Results}
To validate this approach, we perform simulations of a bulk system
of hard spheres and we find that the packing fractions of the
coexisting fluid and face-centered-cubic (fcc) solid phase are
given by $\eta_f=0.4915\pm 0.0005$ and $\eta_s=0.5428\pm0.0005$,
respectively. The pressure and the chemical potential at
coexistence are $\beta P\sigma^3=11.57\pm0.10$ and $\beta\mu=16.08
\pm 0.10$. These results are in good agreement with earlier
results \cite{Hoover1968,Davidchack1998}. Furthermore, to validate
the approach for confined systems, we determine at bulk
coexistence the wall-fluid interfacial tension $\beta\gamma_{\rm
wf}\sigma^2=1.990\pm 0.007$, and the wall-solid interfacial
tension for the (111) and (100) planes of the fcc phase, $\beta
\gamma^{111}_{\rm ws}\sigma^2=1.457 \pm 0.018$ and
$\beta\gamma^{100}_{\rm ws}\sigma^2=2.106\pm0.021$. Our results
are in agreement with previous simulations \cite{Heni1999},  but
the statistical error is one order of magnitude smaller due to our
new thermodynamic integration path.

\begin{figure}
  \begin{center} \begin{indented} 
\item[]
     \includegraphics[width=5cm]{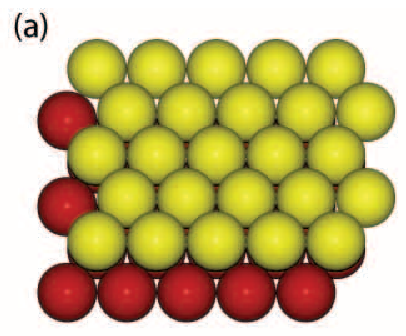}
     \includegraphics[width=5cm]{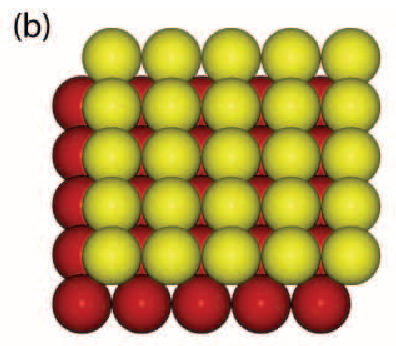}\\
      \includegraphics[width=5cm]{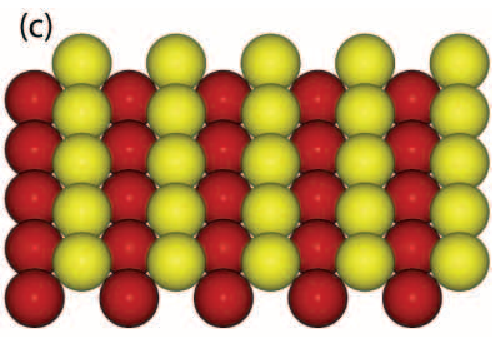}
      \includegraphics[width=5cm]{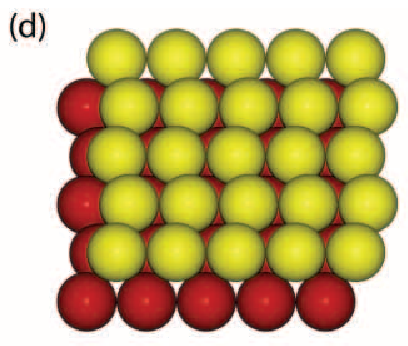}\\
        \includegraphics[width=5cm]{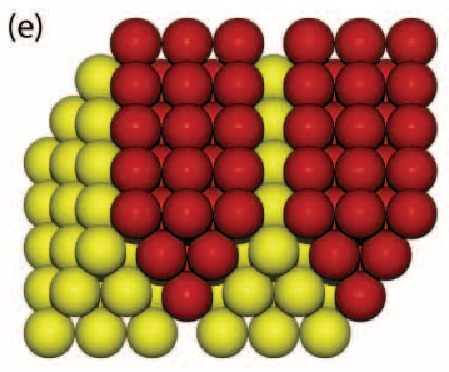}
   \includegraphics[width=5cm]{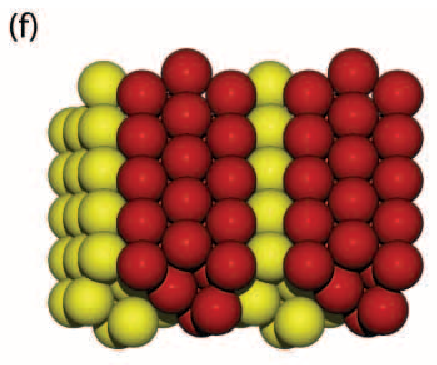}  \\
       \includegraphics[width=5cm]{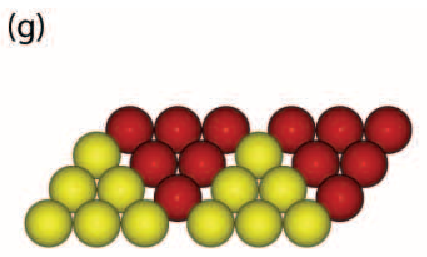}
      \includegraphics[width=5cm]{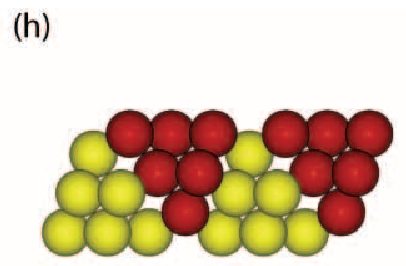}\\
      \end{indented}
       \caption{(Color online) Stable solid structures of confined hard spheres. (a) The triangular phase $2\trr$ (b) The
       square phase $2\sqq$ (c) The buckling phase $2\bk$ (d) The rhombic phase $2\rh$ (e),(g) The prism phase with square symmetry $3\ps$ (f),(h) The prism phase with triangular symmetry $3\ptt$.   In (a)-(f)  the point of view is at an angle of $30^\circ$ to the z direction. In (g),(h) the point of view is at an angle of  $90^\circ$. Different shades (colors) indicate particles in different planes ((a)-(d)) or particles belonging to different prism structures ((e)-(h)).}
    \label{sn:rh}
  \end{center}
\end{figure}

Employing this approach we determine the phase behavior of
confined hard spheres  for plate separations $1 < H/\sigma \leq
5$.
Fig.~\ref{F:phd1} displays the full phase diagram based on free
energy calculations in the $H-\eta$ representation. The white
regions of the phase diagram denote the stable one-phase regions.
The (yellow) shaded regions indicate coexistence between fluid and
solid or two solid phases, and the dotted region is forbidden as
it exceeds the maximum packing fraction of confined hard spheres.
At low densities, we observe a stable fluid phase followed by  a
fluid-solid transition upon increasing the density. The
oscillations in the freezing and melting lines reflect the
(in)commensurability of the crystal structures with the available
space between the walls.  For the crystal phases, we follow the
convention introduced by Pieranski
 \cite{Pieranski1983}, where $n\trr$ denotes a stack
of $n$ triangular layers, and $n\sqq$ a stack of $n$ square layers.
For $H/\sigma \rightarrow 1$, the stable crystal phase consists of
a single triangular layer $1\trr$, which packs more efficiently
than the square layer. As the gap between the plates increases,
crystal slabs with triangular (Fig.~\ref{sn:rh}(a)) and square
packings (Fig.~\ref{sn:rh}(b)) are alternately stable. We find the
characteristic sequence $\cdots n \trr \rightarrow (n+1)\sqq
\rightarrow (n+1) \trr$, which consists of an $n \trr  \rightarrow
(n+1) \sqq$ transformation where both the number of layers and the
symmetry change followed by an $(n+1) \sqq  \rightarrow(n+1) \trr$
transformation where only the symmetry changes.  This sequence is
driven by a competition of a smaller height of $n$ square layers
compared to $n$ triangular layers and a more efficient packing of
triangular layers w.r.t. square layers. When the available gap is
larger than required for the $n\trr$ structure, but smaller than
for $(n+1)\sqq$, intermediate structures may become stable. Similar
arguments can be used for the intervention of intermediate
structures in the $(n+1)\sqq \rightarrow (n+1)\trr$ transformation.
Especially at high packing fractions, the spheres can increase
their packing by adopting interpolating structures. In
Fig.~\ref{F:phd1} we report the boundaries of the interpolating
regions $In$. Each region represents one or more interpolating
structures, that are listed in Tab.~\ref{tab:list}, according to
the standard notation. Within the resolution of our simulations,
it is difficult to draw the phase boundaries of all the
intermediate structures in $In$, but in Tab.~\ref{tab:list} the
{\em thermodynamically stable} structures are listed in the order
they appear upon increasing $H$ and $\eta$. We also compare our
sequence of structures with the experimental one
\cite{Fontecha2005}. The experiments considered charged particles,
but we do not expect that the soft repulsion has a strong  effect
on the observed structures at high densities. The agreement is
excellent at small plate separations. The buckling phase $2\bk$
(Fig.~\ref{sn:rh}(c)) interpolates between $1\trr$ and $2\sqq$. In
the $2\bk$, the $1\trr$ is split into two sublayers consisting of
rows that are displaced in height and which can transform smoothly
into $2\sqq$ upon increasing the gap. The rhombic phase $2\rh$
(Fig.~\ref{sn:rh}(d)) is found between $2\sqq$ and $2\trr$. The
rhombic phase  is also stable between $n\sqq$ and $n\trr$ for
$n\leq5$, but not in the whole region. In addition, we find that
at higher $n$ the interpolating structures are mainly prism
phases.  In agreement with experiments, we  find two types of
prisms, one with a square base $n\ps$ (Fig.~\ref{sn:rh}(e)), and
one with a triangular base $n\ptt$ (Fig.~\ref{sn:rh}(f)), where $n$
indicates the number of particles in the prism base. As shown in
Fig.~\ref{sn:rh}(g),(h), these structures display large gaps as a
result of periodically repeated stacking faults in the packing
which, nevertheless, allow particles to pack more efficiently,
than a phase consisting of parallel planes of particles.  For
$n>3$, differences between simulations and experiments emerge. We
find that the stability region of interpolating structures between
$n\trr$, and $(n+1)\sqq$ decreases for larger $H$, becoming
invisible on the scale of Fig.~\ref{F:phd1} for  $I9=5\trr
\rightarrow 6\sqq$. On the other hand, the region of stability of
the interpolating structures between $n\sqq$ and $n\trr$ increases
while increasing the wall separation, becoming stable also at low
packing fractions for the transitions $I8=5\sqq \rightarrow 5\trr$,
and possibly $I10=6\sqq \rightarrow 6\trr$. We also note that the
solid-solid transitions are first-order with a clear density jump
at low $\eta$, but they get weaker (and maybe even continuous)
upon approaching the maximum packing limit. In addition, the
rhombic and buckling phases are highly degenerate as we find
zig-zag and linear buckling or rhombic phases, and a combination
of those.

\begin{figure}
  \begin{center} 
  \begin{indented} 
\item[]  \includegraphics[width=10cm]{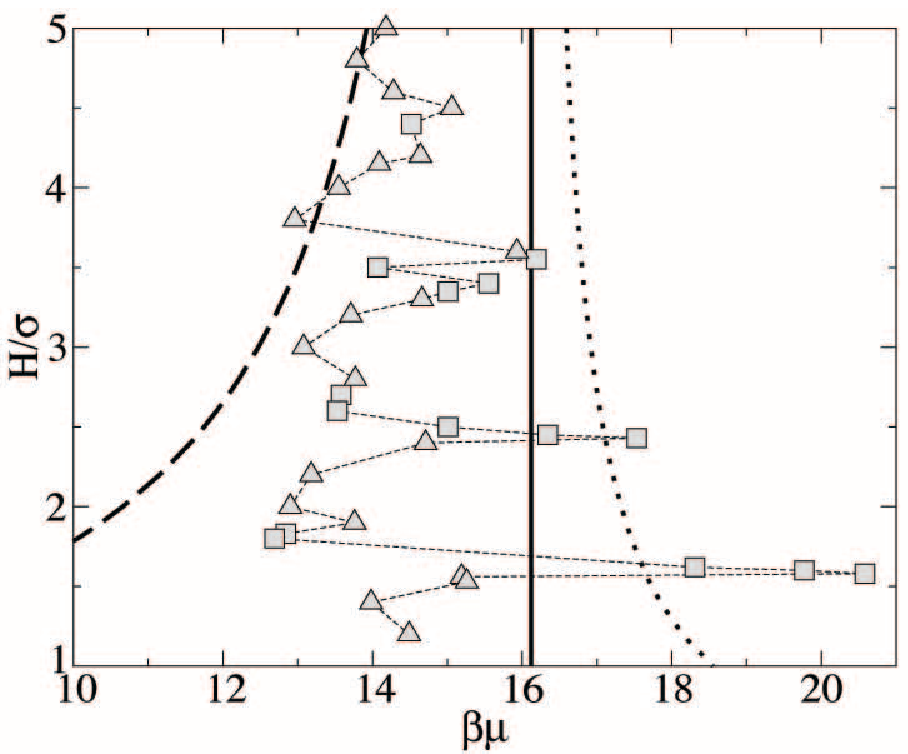}
\end{indented}
    \caption{Chemical potential $\beta \mu$ at fluid-solid coexistence, for different wall separations $H/\sigma$.
    The symbols are the simulation results for the triangular ($\trr$) and square structures ($\sqq$). The thin dashed
    line is a guide to the eye. The thick continuous line indicate the value of the bulk freezing chemical potential
    $\beta \mu=16.08$. The thick dashed and dotted curves are the prediction of the Kelvin equation for the $111$, and $100$
    planes parallel to the walls, respectively. }
    \label{F:phd2}
  \end{center}
\end{figure}

We now turn our attention to the fluid-solid transition. In
Fig.~\ref{F:phd2}, we plot the chemical potential $\beta \mu^{\rm
cap}$ at the freezing transition of the confined system as a
function of $H$. The freezing for crystal slabs with a triangular
symmetry are denoted by triangles, while the square symmetry is
displayed by squares. We find strong oscillations in the chemical
potential reminiscent to the (in)commensurability of the crystal
structures with plate separation. The highest values for $\beta
\mu^{\rm cap}$ are reached at the transition region $n\trr
\rightarrow (n+1)\sqq$, corresponding to plate separations where
both structures are incommensurate and hence unfavorable. In this
regime, $\beta \mu^{\rm cap}$ can reach values that are higher
than the bulk freezing chemical potential $\beta \mu^{\rm bulk}$
(the black vertical line in Fig.~\ref{F:phd2}), corresponding to
capillary melting, while the freezing transitions with $\beta
\mu^{\rm cap}$ lower than the bulk value correspond to capillary
freezing. Hence, we find a reentrant capillary freezing/melting
behavior for wall separations $1<H/\sigma<3.5$. In addition, we
compare our results with the predictions of the Kelvin
equation\cite{Rowlinson2002}: $\beta \mu^{\rm cap} = \beta
\mu^{\rm bulk} - \pi\sigma^3/3H
(\gamma_{wf}-\gamma_{ws})/(\eta_s-\eta_f)$
using the parameters determined in our simulations. The thick
dashed line in  Fig.~\ref{F:phd2} is the prediction of the Kelvin
equation  for the $(111)$ crystal plane (triangular order) at the
walls, while the dotted line is that for the $(100)$ plane (square
order). The Kelvin equation predicts capillary freezing for the
triangular structure and capillary melting for the square
structures.  The Kelvin equation predictions are in reasonable
agreement  with our simulations for triangular order for wall
separation as small as $H/\sigma \sim 4$, but deviates for smaller
$H$, while the prediction for the square structure is in agreement
only at very small $H$. It is surprising to find qualitative
agreement at small $H$ since the Kelvin equation is valid in the
limit $H/\sigma \rightarrow \infty$.
\section{Conclusion}
In summary, we have calculated the equilibrium phase diagram of
confined hard spheres using free energy calculations with a novel
integration path.  The high density sequence of structures is in
good agreement with experimental results. We find that the prism
phases are thermodynamically stable also at lower densities, and
this work will, hopefully, stimulate further experimental
investigations, for a quantitative comparison at intermediate
packing fractions. In addition, our results show an intriguing
sequence of melting and freezing transitions upon increasing the
distance between the walls of a slit which is in contact with a
bulk reservoir. The mechanical behavior is therefore very
sensitive on the degree of confinement, and the knowledge of the
phase diagram can help the understanding and fabrication of new
materials. The transition from confined to bulk behavior, and the
interface between different solid structures (studied in lower
dimensions in Ref.~\cite{Chaudhuri2004}) represent interesting
directions for future investigations.

\ack
We thank M. Schmidt for inspiring discussions. This work is
part of the research program of the {\em Stichting voor
Fundamenteel Onderzoek der Materie} (FOM), that is financially
supported by the {\em Nederlandse Organisatie voor
Wetenschappelijk Onderzoek} (NWO).  We thank the Dutch National
Computer Facilities foundation for granting access to TERAS and
ASTER supercomputers.

\section*{References}
\bibliographystyle{unsrtd}

\begin{thebibliography}{99}
\bibitem{Hoover1968}
W.G. Hoover and F.M. Ree, J. Chem. Phys. {\bf 49}, 3609 (1968).
\bibitem{Pusey1986}
P. N. Pusey and W. van Megen, Nature (London) {\bf 320}, 340
(1986).
\bibitem{Dijkstra2004}
M. Dijkstra, Phys. Rev. Lett. {\bf 93}, 108303 (2004); D.J.
Courtemanche and F. van Swol, Phys. Rev. Lett. {\bf 69}, 2078
(1992).
\bibitem{Pieranski1983}
P. Pieranski and L. Strzelecki and B. Pansu, Phys. Rev. Lett. {\bf
50}, 900 (1983).
\bibitem{Neser1997}
S. Neser {\em et al.}, Phys. Rev. Lett. {\bf 79}, 2348 (1997).
\bibitem{Fontecha2005}
A.B. Fontecha {\em et al.}, J. Phys.: Cond. Matt. {\bf 17}, S2779
(2005).
\bibitem{Cohen2004}
I. Cohen {\em et al.}, Phys. Rev. Lett. {\bf 93}, (2004).
\bibitem{Schmidt1996}
M. Schmidt and H. L\"owen, Phys. Rev. Lett. {\bf 76}, 4552 (1996);
Phys. Rev. E {\bf 55}, 7228 (1997).
\bibitem{Zangi2000}
R. Zangi and S.A. Rice, Phys. Rev. E {\bf 61}, 660 (2000); ibid
{\bf 61}, 671 (2000).
\bibitem{Messina2003}
R. Messina and H. L\"owen, Phys. Rev. Lett. {\bf 91}, 146101
(2003);  Phys. Rev. E {\bf 73}, 011405 (2006).
\bibitem{Gao1997} J.
Gao {\em et al.}, Phys. Rev. Lett. {\bf 79}, 705 (1997).
\bibitem{Klein1995} J.
Klein and E. Kumacheva, Science. {\bf 269}, 816 (1995).
\bibitem{Ghatak2001} C. Ghatak and K. G. Ayappa, Phys. Rev. E. {\bf 64}, 051507 (2001).
\bibitem{Ciach2004} A. Ciach, Progr. Colloid Polym. Sci. {\bf 129}, 40 (2004).
\bibitem{book}
J.H. Conway and N.J.A. Sloane, {\it Sphere Packings, Lattices and
Groups} (Springer, New York, 1993); T. Aste and D. Weaire, {\it
The Pursuit of Perfect Packing} (IOP, Bristol, 2000).
\bibitem{Frenkel2002}
D. Frenkel and B. Smit, {\it Understanding {M}olecular
{S}imulation 2nd edition} (Academic Press, 2002).
\bibitem{Fortini2005a}
A. Fortini {\em et al.}, Phys. Rev. E {\bf 71}, 051403 (2005).
\bibitem{evans}
R. Evans and U. Marini Bettolo Marconi, J. Chem. Phys. {\bf 86},
7138 (1987).
\bibitem{Davidchack1998}
R.L. Davidchack and B.B. Laird, J. Chem. Phys. {\bf 108}, 9452
(1998).
\bibitem{Heni1999}
M. Heni and H. L\"owen, Phys. Rev. E {\bf 60}, 7057 (1999).
\bibitem{Rowlinson2002}
J.S. Rowlinson and B. Widom, {\it Molecular {T}heory of
{C}apillarity} (Dover, New York, 2002).


\bibitem{Chaudhuri2004} D.
Chaudhuri and S. Sengupta. Phys. Rev. Lett. {\bf 93}, 115702 (
2004).
\end{thebibliography}

\end{document}